\documentclass[runningheads]{llncs}
\usepackage{graphicx}
\usepackage{wrapfig}
\usepackage{tikz}
\usepackage{comment}
\usepackage{amsmath,amssymb} 
\usepackage{color}
\usepackage[accsupp]{axessibility}  

\usepackage{kotex} 
\usepackage{booktabs}
\usepackage{url}
\usepackage[ruled,vlined]{algorithm2e}
\usepackage{amsmath}
\definecolor{commentcolor}{RGB}{110,154,155}   

\usepackage{subcaption}
\usepackage{floatrow}
\newfloatcommand{capbtabbox}{table}[][\FBwidth]
\usepackage{hyperref}

\begin{document}
\pagestyle{headings}
\mainmatter
\def\ECCVSubNumber{25}  

\title{Learning Multiple Probabilistic Degradation Generators for Unsupervised Real World Image Super Resolution}

\titlerunning{ECCV-22 submission ID \ECCVSubNumber} 
\authorrunning{ECCV-22 submission ID \ECCVSubNumber} 
\author{Anonymous ECCV submission}
\institute{Paper ID \ECCVSubNumber}

\titlerunning{Learning Multiple Probabilistic Degradation Generator}
%
\author{Sangyun Lee\thanks{Author did this work during the intership at SI Analytics.}\inst{1} \and
Sewoong Ahn\inst{2} \and
Kwangjin Yoon\thanks{Corresponding author.}\inst{2}}
\authorrunning{Lee et al.}
%
\institute{Soongsil University \and
SI Analytics \\
\email{ml.swlee@gmail.com \tt \{anse3832, yoon28\}@si-analytics.ai}}

\def\QED{~\rule[-1pt]{5pt}{5pt}\par\medskip}

\newcommand{\term}[1] {\emph{#1}}
\newcommand{\reg} {\xmath{\beta}}

\newcommand{\fref}[1] {Fig.~\ref{#1}\xspace}
\newcommand{\tref}[1]{Table~\ref{#1}}

\newcommand{\diag}{\operatorname{diag}}
\newcommand{\tr}{\operatorname{tr}}
\newcommand{\st}{\operatorname{s.t.}}
\newcommand{\range}{\operatorname{Range}}
\newcommand{\rank}{\operatorname{rank}}
\renewcommand{\dim}{\operatorname{dim}}
\newcommand{\pri}{\text{pri}}
\newcommand{\dual}{\text{dual}}

\newcommand{\mathbfat}[1] {\begin{bmatrix} #1 \end{bmatrix}} 

\newcommand{\xmath}[1] {\ensuremath{#1}\xspace}
\newcommand{\A}{\mathbf{A}}
\newcommand{\B}{\mathbf{B}}
\newcommand{\C}{\mathbf{C}}
\newcommand{\D}{\mathbf{D}}
\newcommand{\E}{\mathbf{E}}
\newcommand{\F}{\mathbf{F}}
\newcommand{\I}{\mathbf{I}}
\newcommand{\R}{\mathbf{R}}
\newcommand{\T}{\mathbf{T}}
\newcommand{\U}{\mathbf{U}}
\newcommand{\V}{\mathbf{V}}
\newcommand{\W}{\mathbf{W}}
\newcommand{\X}{\mathbf{X}}
\newcommand{\Y}{\mathbf{Y}}
\newcommand{\Z}{\mathbf{Z}}
\renewcommand{\P}{\mathbf{P}} 
\renewcommand{\H}{\mathbf{H}}
\newcommand{\G}{\mathbf{G}}

\newcommand{\x}{\mathbf{x}}
\newcommand{\f}{\mathbf{f}}
\newcommand{\xh}{\xmath{\hat{\x}}}
\newcommand{\y}{\mathbf{y}}
\newcommand{\w}{\mathbf{w}}
\newcommand{\z}{\mathbf{z}}
\newcommand{\zh}{\xmath{\hat{\z}}}
\newcommand{\1}{\mathbf{1}} 
\newcommand{\0}{\mathbf{0}} 
\renewcommand{\Re}{\mathbb{R}}
\newcommand{\ie}{\text{i.e.}}
\newcommand{\eg}{\text{e.g.}}

\newcommand{\bLam}{\mathbf{\Lambda}}

\newcommand{\norm}[1] {\xmath{\left\| #1 \right\|}}
\newcommand{\normi}[1] {\xmath{\left\| #1 \right\|_1}}
\newcommand{\normii}[1] {\xmath{\left\| #1 \right\|_2}}


\newcommand{\Cm}{{\mathcal C}}
\newcommand{\Ab}{{\mathbf A}}
\newcommand{\Bb}{{\mathbf B}}
\newcommand{\Cb}{{\mathbf{C}}}
\newcommand{\Db}{{\mathbf D}}
\newcommand{\Eb}{{\mathbf E}}
\newcommand{\Fb}{{\mathbf F}}
\newcommand{\Gb}{{\mathbf G}}
\newcommand{\Hb}{{\mathbf H}}
\newcommand{\Ib}{{\mathbf I}}
\newcommand{\Jb}{{\mathbf J}}
\newcommand{\Kb}{{\mathbf K}}
\newcommand{\Lb}{{\mathbf L}}
\newcommand{\Mb}{{\mathbf M}}
\newcommand{\Nb}{{\mathbf N}}
\newcommand{\Ob}{{\mathbf O}}
\newcommand{\Pb}{{\mathbf P}}
\newcommand{\Qb}{{\mathbf Q}}
\newcommand{\Rb}{{\mathbf R}}
\newcommand{\Sb}{{\mathbf S}}
\newcommand{\Tb}{{\mathbf T}}
\newcommand{\Ub}{{\mathbf U}}
\newcommand{\Vb}{{\mathbf V}}
\newcommand{\Wb}{{\mathbf W}}
\newcommand{\Xb}{{\mathbf X}}
\newcommand{\Yb}{{\mathbf Y}}
\newcommand{\Zb}{{\mathbf Z}}

\newcommand{\ab}{{\mathbf a}}
\newcommand{\bb}{{\mathbf b}}
\newcommand{\cb}{{\mathbf c}}
\newcommand{\db}{{\mathbf d}}
\newcommand{\eb}{{\mathbf e}}
\newcommand{\fb}{{\mathbf f}}
\newcommand{\gb}{{\mathbf g}}
\newcommand{\hb}{{\mathbf h}}
\newcommand{\ib}{{\mathbf i}}
\newcommand{\jb}{{\mathbf j}}
\newcommand{\kb}{{\mathbf k}}
\newcommand{\lb}{{\mathbf l}}
\newcommand{\mb}{{\mathbf m}}
\newcommand{\nb}{{\mathbf n}}
\newcommand{\ob}{{\mathbf o}}
\newcommand{\pb}{{\mathbf p}}
\newcommand{\qb}{{\mathbf q}}
\newcommand{\rb}{{\mathbf r}}
\renewcommand{\sb}{{\mathbf s}}
\newcommand{\tb}{{\mathbf t}}
\newcommand{\ub}{{\mathbf u}}
\newcommand{\vb}{{\mathbf v}}
\newcommand{\wb}{{\mathbf w}}
\newcommand{\xb}{{\mathbf x}}
\newcommand{\yb}{{\mathbf y}}
\newcommand{\zb}{{\mathbf z}}
\newcommand{\Rc}{\mathcal{R}}
\newcommand{\mbf}{\mathbf}
\newcommand{\bs}{\boldsymbol}
\newcommand{\Hs}{\mathscr{H}}
\newcommand{\Xs}{\mathscr{X}}
\newcommand{\Hc}{\mathcal{H}}
\newcommand{\Gc}{\mathcal{G}}
\newcommand{\Ac}{\mathcal{A}}
\newcommand{\Bc}{\mathcal{B}}
\newcommand{\Cc}{\mathcal{C}}
\newcommand{\Lc}{\mathcal{L}}
\newcommand{\Nc}{\mathcal{N}}
\newcommand{\Qc}{\mathcal{Q}}
\newcommand{\Tc}{\mathcal{T}}
\newcommand{\Xc}{\mathcal{X}}
\newcommand{\Sc}{\mathcal{S}}
\newcommand{\Yc}{\mathcal{Y}}
\newcommand{\Zc}{\mathcal{Z}}
\newcommand{\Vc}{\mathcal{V}}
\newcommand{\Wc}{\mathcal{W}}

\newcommand{\Phib}{{\boldsymbol {\Phi}}}
\newcommand{\Psib}{{\boldsymbol {\Psi}}}
\newcommand{\Thetab}{{\boldsymbol {\Theta}}}
\newcommand{\Sigmab}{{\boldsymbol {\Sigma}}}
\newcommand{\Rd}{{\mathbb R}}
\newcommand{\Cd}{{\mathbb C}}
\newcommand{\Nd}{{\mathbb N}}
\newcommand{\Id}{{\mathbb I}}
\newcommand{\Fd}{{\mathbb F}}
\newcommand{\deltab}{{\boldsymbol{\delta}}}
\newcommand{\mub}{{\boldsymbol{\mu}}}
\newcommand{\nub}{{\boldsymbol{\nu}}}
\newcommand{\phib}{{\boldsymbol{\phi}}}
\newcommand{\psib}{{\boldsymbol{\psi}}}
\newcommand{\rhob}{{\boldsymbol {\rho}}}
\newcommand{\alphab}{{\boldsymbol {\alpha}}}
\newcommand{\Ybc}{{\boldsymbol{\mathcal Y}}}
\newcommand{\Sbc}{{\boldsymbol{\mathcal S}}}
\newcommand{\Qbc}{{\boldsymbol{\mathcal Q}}}
\newcommand{\Zbc}{{\boldsymbol{\mathcal Z}}}
\newcommand{\Abc}{{\boldsymbol{\cal A}}}
\newcommand{\Ebc}{{\boldsymbol{\cal E}}}
\newcommand{\Cbc}{{\boldsymbol{\cal C}}}
\newcommand{\Rbc}{{\boldsymbol{\mathcal R}}}
\newcommand{\thetab}{{\boldsymbol {\theta}}}
\newcommand{\chib}{{\boldsymbol {\chi}}}
\newcommand{\xib}{{\boldsymbol {\xi}}}
\newcommand{\Hbc}{{\boldsymbol{\mathcal H}}}
\newcommand{\Xbc}{{\boldsymbol{\mathcal X}}}
\newcommand{\Kbc}{{\boldsymbol{\mathcal K}}}
\newcommand{\Mbc}{{\boldsymbol{\mathcal M}}}
\newcommand{\Kc}{{{\mathcal K}}}
\newcommand{\Pc}{{{\mathcal P}}}
\newcommand{\Ec}{{{\mathcal E}}}
\newcommand{\Dc}{{{\mathcal D}}}

\newcommand{\Ed}{{{\mathbb E}}}
\newcommand{\Zd}{\mathbb{Z}}
\newcommand{\hank}{\boldsymbol{\mathbb{H}}}
\newcommand{\cir}{\mathscr{C}}
\newcommand{\conv}{\boldsymbol{\mathbb{C}}}
\newcommand{\Tconv}{\mathcal{T}}
\newcommand{\zerob}{\mathbf{0}}
\newcommand{\hbk}{\boldsymbol{\mathfrak{h}}}
\newcommand{\Null}{\textsc{Nul}}
\newcommand{\Ran}{\textsc{Ran}}
\newcommand{\id}{\mathrm{id}}

\newcommand{\Md}{\mathbb{M}}
\newcommand{\Kd}{\mathbb{K}}
\newcommand{\Jd}{\mathbb{J}}
\newcommand{\Td}{\mathbb{T}}
\newcommand{\imsize}{6.5}
\newcommand{\imsizer}{6.2}
\newcommand{\imsizeh}{3.2}
\newcommand{\imsizes}{4}

\newcommand{\beq}{\begin{equation}}
\newcommand{\eeq}{\end{equation}}
\newcommand{\beqa}{\begin{eqnarray}}
\newcommand{\eeqa}{\end{eqnarray}}
\newcommand{\Fc}{{\mathcal F}}

\newcommand{\vect}{\textsc{Vec}}
\newcommand{\lambdab}{\boldsymbol{\lambda}}
\newcommand{\Lambdab}{\boldsymbol{\Lambda}}
\maketitle

\begin{abstract}
  Unsupervised real world super resolution (USR) aims to restore high-resolution (HR) images given low-resolution (LR) inputs, and its difficulty stems from the absence of paired dataset.
  One of the most common approaches is synthesizing noisy LR images using GANs (i.e., degradation generators) and utilizing a synthetic dataset to train the model in a supervised manner.
  Although the goal of training the degradation generator is to approximate the distribution of LR images given a HR image, previous works have heavily relied on the unrealistic assumption that the conditional distribution is a delta function and learned the deterministic mapping from the HR image to a LR image.
  In this paper, we show that we can improve the performance of USR models by relaxing the assumption and propose to train the probabilistic degradation generator.
  Our probabilistic degradation generator can be viewed as a deep hierarchical latent variable model and is more suitable for modeling the complex conditional distribution. 
  We also reveal the notable connection with the noise injection of StyleGAN.
  Furthermore, we train multiple degradation generators to improve the mode coverage and apply \textit{collaborative learning} for ease of training.
  We outperform several baselines on benchmark datasets in terms of PSNR and SSIM and demonstrate the robustness of our method on unseen data distribution.
  Code is available at \href{https://github.com/sangyun884/MSSR}{https://github.com/sangyun884/MSSR}.

\end{abstract}

\section{Introduction}
\label{sec:intro}

\begin{figure}
\centering
\begin{tabular}{cc}
\includegraphics[height=3.5cm]{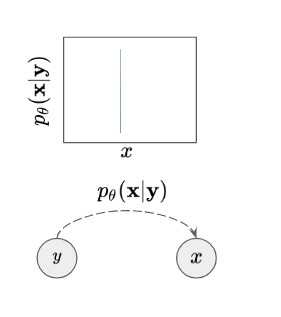}&
\includegraphics[height=3.5cm]{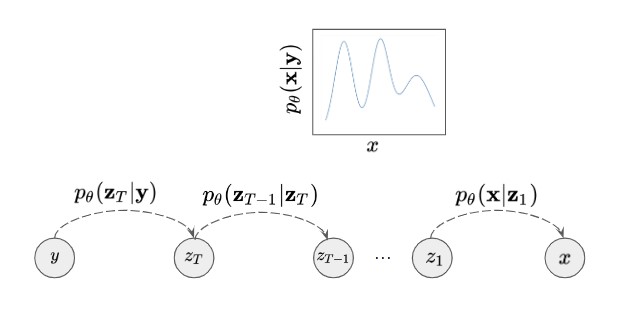}\\
(a)  &(b) 
\end{tabular}
\caption{The most crucial step of USR is to learn the degradation process in which HR image \(\y\) transforms into LR image \(\x\). (a) : Previous GAN-based studies model the $p(\x | \y)$ as a delta function. (b) : We design the probabilistic generator with multiple latent variables to model the flexible distribution.}
\label{fig:circles}
\end{figure}

Unsupervised real world super resolution (USR) aims to restore high-resolution (HR) images given low-resolution (LR) observations, and its difficulty stems from the absence of paired dataset.
While recent progress of deep convolutional neural network-based approaches has shown remarkable results on the bicubic downsampled dataset~\cite{dong2015image,kim2016accurate,lim2017enhanced,shi2016real,zhang2018image}, they generalize poorly in the real world as they do not consider the complex degradation process.

In real world, the degradation process is generally unknown and only implicitly represented through the corrupted observations.
Since it is difficult to collect paired data consisting of LR images and their intact HR representations, several studies~\cite{Bulat_2018_ECCV,lugmayr2019unsupervised,fritsche2019frequency} have focused on generating realistic synthetic paired data using conditional generative models, primarily generative adversarial networks (GANs)~\cite{goodfellow2014generative}.
In contrast to the approaches dubbed blind SR~\cite{wang2021real,wang2021unsupervised}, GAN-based approaches are effective in the real world as they do not assume any functional form of the degradation process, which can actually be a complex and nonlinear. 

However, previous GAN-based approaches have heavily relied on the unrealistic assumption that the degradation process is deterministic and learned the deterministic mapping from a HR image to a LR image.
Since they fail to capture the complex distribution of LR images due to their deterministic nature (see Fig.~\ref{fig:circles} (a)), the generalization capability of a SR model trained on the pseudo-paired dataset is sub-optimal.

In this paper, we show that we can improve the performance of USR models by relaxing the assumption and propose to train the probabilistic degradation generator to better approximate the stochastic degradation process.
As shown in Fig.~\ref{fig:circles}, our probabilistic degradation generator can be viewed as a deep hierarchical latent variable model and is more suitable for modeling the complex conditional distribution.
Our method is conceptually simple and extremely easy to implement: add a Gaussian noise multiplied by the learned standard deviation to intermediate feature maps of the degradation generator.
We also reveal that the noise injection of StyleGAN~\cite{karras2019style} can be viewed as a special case of our method. 

Furthermore, we train multiple degradation generators to improve the mode coverage and apply \textit{collaborative learning} for ease of training.
In collaborative learning, multiple SR models teach with one another, mutually improving their generalization capabilities.
We show that collaborative learning facilitates the training procedure.

In addition, we outperform several baselines on benchmark datasets in PSNR and SSIM and demonstrate the robustness of our method on unseen data distribution. Our contributions are summarized as follows:
\begin{itemize}
\item[--] We propose the probabilistic degradation generator to model the stochastic degradation process and reveal the connection with the noise injection of StyleGAN.
\item[--] We train multiple degradation generators to improve the mode coverage and apply a collaborative learning strategy to facilitate the training procedure.
\item[--] Our model outperforms several baselines on benchmark datasets and shows a strong out-of-domain generalization ability.
\end{itemize}

\begin{figure*}
	\centering

	\begin{tabular}{cc}
	\includegraphics[height=5cm]{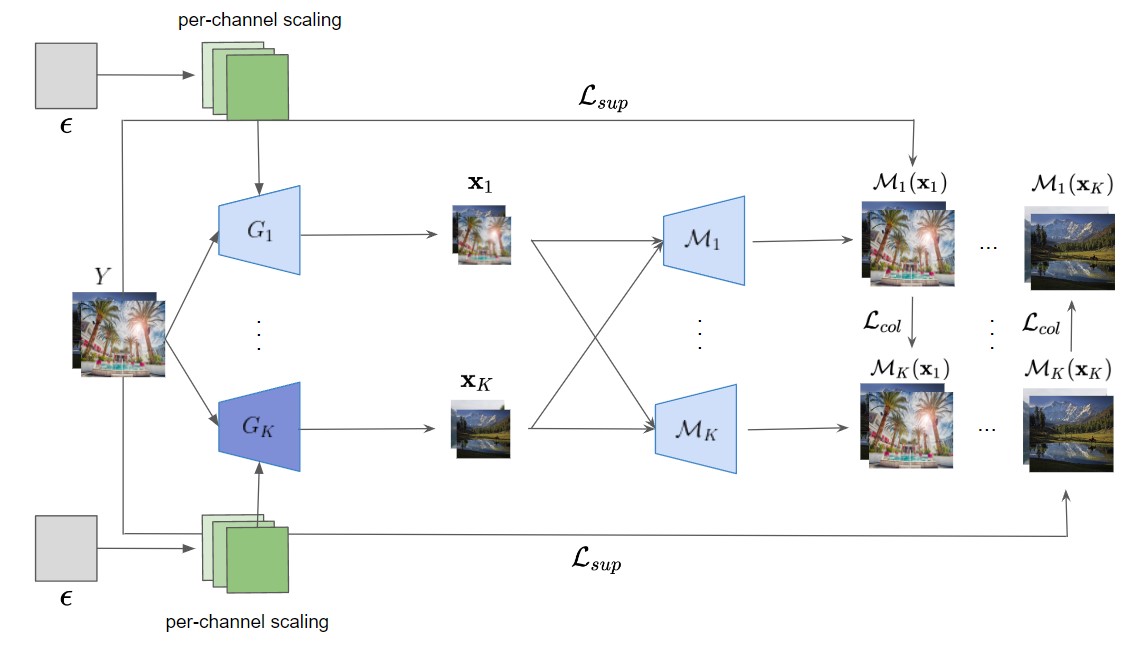}&
	\includegraphics[height=5cm]{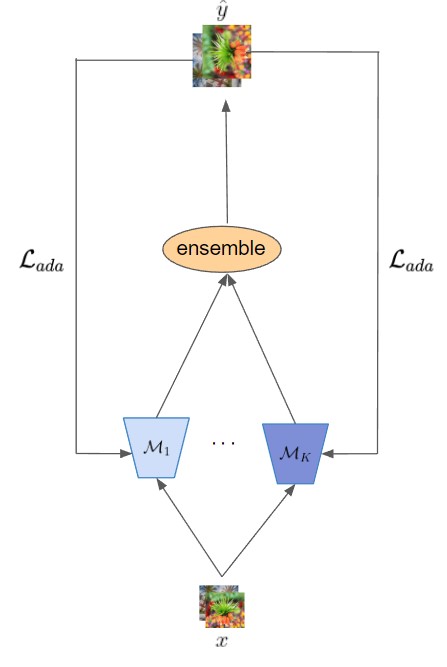} \\
	(a) & (b)

	\end{tabular}
	\caption{A schematic of collaborative learning. (a) : Collaborative learning on the synthetic LR domain using multiple probabilistic degradation generators. (b) : Collaborative learning on the real LR dataset using pseudo-labels. We alternatively train (a) and (b).}
	\label{fig:overview}
\end{figure*}

\section{Related Works}
\subsection{GAN-based Unsupervised Real-World Super Resolution}
Although previous SR models~\cite{dong2015image,kim2016accurate,lim2017enhanced,shi2016real,zhang2018image} achieved remarkable results on bicubic downsampled data, they perform poorly in the real world, where the degradation process is complex and unknown. Several studies~\cite{Bulat_2018_ECCV,lugmayr2019unsupervised,fritsche2019frequency} have attempted to model the degradation process using GANs. They use conditional GANs to generate LR images given HR images and train the SR model on the generated pseudo-paired data.
However, they use a deterministic degradation model and fail to approximate the stochastic degradation process, to which most of the real world scenarios belong.

\subsection{Stochastic Degradation Operations}
There have been studies called blind SR that take account of the stochastic degradation process~\cite{wang2021real,wang2021unsupervised}. However, they assume the limited type of degradation operation that consists of blur kernel and noise from the pre-defined distribution.
Recently, DeFlow~\cite{wolf2021deflow} tackled this problem by learning unpaired learning with normalizing flow.
In this paper, we propose a method that is more straightforward and easy to plug into any existing degradation network to enhance the expressiveness.

\subsection{Noise Injection to Feature Space}
Our work is closely related to the approaches that inject the noise into the feature space of the generator.
For example, Karras et al. proposed a noise injection module that helps a model generate the pseudo-random details of images~\cite{Karras_2019_CVPR}.
Figuring out the effectiveness of the noise injection is still an open problem~\cite{feng2021understanding}, and they may be related to modern hierarchical generative models~\cite{vahdat2020nvae,kingma2016improved,ho2020denoising,song2019generative} as the reparameterization trick can be viewed as injecting noises to intermediate states.
We consider our probabilistic degradation generator as a hierarchical generative model with multiple latent variables and will subsequently reveal the connection with the noise injection of StyleGAN.

\section{Proposed Methods}
\label{sec:others}

\subsection{Probabilistic Degradation Generator}

\begin{figure}
	\centering
	\includegraphics[width=8cm]{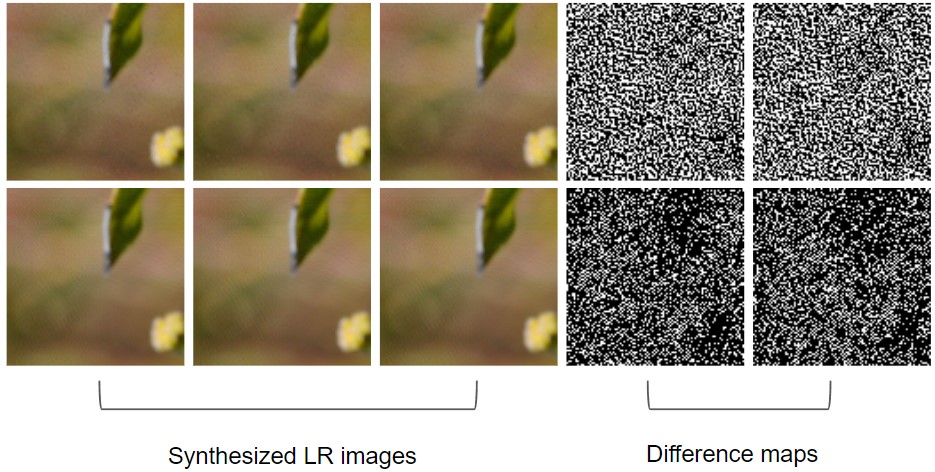}
	\caption{LR images (first three columns) generated by probabilistic degradation generators and difference maps (last two columns). The images of the first row are generated by probabilistic DeResnet~\protect\cite{wei2020unsupervised}, and the images of the second row are generated by probabilistic HAN~\protect\cite{niu2020single}. The difference maps are calculated as the pixel-wise L1 distance between the first and second images and the first and third images, respectively. The proposed probabilistic degradation generator can synthesize multiple LR images given a single HR observation.  }
	\label{fig:compare_sources}
\end{figure}
From a probabilistic perspective, training the degradation generator is to estimate the parameter \(\theta\) that satisfies \(p_\theta(\x|\y) \approx p(\x|\y)\), where \(\x \sim p(\x)\) and \(\y \sim p(\y)\) are a LR and HR image.
Our aim is to generalize the deterministic degradation generators in the previous studies to better approximate the stochastic degradation process in the real world.
With \(T\) latent variables \(\z_1, \z_2, ..., \z_T\), the probabilistic degradation generator is defined as:

\begin{equation}
p_\theta(\x|\y) = \int p_\theta(\x|\z_1)p_\theta(\z_T|\y) \prod_{i=2}^T p_\theta(\z_{i-1}|\z_i) d\z_{1:T},
\end{equation} which is the Markov chain that consists of following Gaussian transition:
\begin{equation}
    p_\theta(\z_{i-1}|\z_i) = \mathcal{N}(\z_{i-1};\mathbf{\mu}_\theta(\z_i, i), \mathbf{\Sigma}_\theta(\z_i, i)).
\end{equation}

Note that $p_\theta(\x|\z_1)$ and $p_\theta(\z_T|\y)$ are delta functions. For simplicity and ease of computation, we set $\mathbf{\Sigma}_\theta(\z_i,i) = diag(\mathbf{\sigma}_\theta(\z_i,i))^2$ with $\mathbf{\sigma}_\theta(\cdot)$ being a vector-valued function.
Therefore, we can sample each latent variable in a differentiable manner as
\begin{equation}
    \z_{i-1} = \mathbf{\mu}_\theta(\z_i) + diag(\mathbf{\sigma}_\theta(\z_i,i))\epsilon, \ \ \ \epsilon \sim \mathcal{N}(\mathbf{0}, \Ib),
    \label{eq:noise_injection}
\end{equation}
which can be seen as injecting the noise into the feature vector $\mathbf{\mu}_\theta(\z_i)$.
Note that our model is reduced to the deterministic generator in previous studies when $\mathbf{\sigma}_\theta(\z_i) = \mathbf {0}$.
Since maximum likelihood training is difficult as there is no paired data, we train the probabilistic generator via adversarial learning with cycle-consistency constraint~\cite{zhu2017unpaired}.

\noindent \textbf{Connection with StyleGAN}
When the variances $\mathbf{\Sigma}_\theta$ of latent variables do not depend on $\z_i$ and are shared within the spatial dimension (i.e., the elements in the same channel have the same variance), the second term of the right-hand side of eq.~\ref{eq:noise_injection} is the same as the noise injection in StyleGAN.
In fact, this choice reduces the number of parameters without compromising the performance, and thus we extensively use this setting throughout the manuscript.

\subsection{Training Multiple Degradation Generators}
\label{sec:synthesizing}
GANs are infamous for their mode-dropping trait. Although one can utilize a complicated method to alleviate it, we propose a straightforward way to enhance the mode coverage: to train multiple probabilistic degradation generators \(G_1,...,G_K\).
Note that to maximize the mode coverage, it is crucial for each generator to exclusively cover a certain area of the $p(\x|\y)$ without redundancy.
Depending on the fact that the convergence properties of GANs can vary depending on the training algorithm~\cite{mescheder2018training}, we initially attempted to train each degradation generator with different training schemes and regularizations.
However, we empirically found that the diversity of the LR domain can be efficiently achieved by varying model architectures of the degradation generator.
Fig.~\ref{fig:compare_sources} shows that each generator synthesizes different noise patterns (zoom in for the best view).
We describe detailed information on architectures in sec \ref{sec:imple}.

\begin{figure*}[t]

\begin{floatrow}
{

\begin{tabular}{ccc}

	\includegraphics[width=2cm]{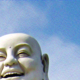}&\includegraphics[width=2cm]{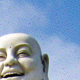}&\includegraphics[width=2cm]{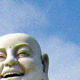}\\
	DeResnet~\cite{wei2020unsupervised} & Probabilistic DeResnet & GT
	\end{tabular}{%
}
}
\capbtabbox{%
  \begin{tabular}{cc} \hline
  Method & LPIPS~\cite{zhang2018unreasonable} \\ \hline
  Gaussian noise  & 0.215 \\
  DeResnet~\cite{wei2020unsupervised} & 0.049 \\
  Probabilistic DeResnet & \bf{0.026} \\ \hline
  \end{tabular}
}{%
}
\end{floatrow}
\caption{Comparison between generated LR images. Since the validation set of NTIRE2020 dataset~\protect\cite{lugmayr2020ntire} consists of paired data, we can measure LPIPS between LR images generated by degradation generators and the ground truth LR image. We also report the result of additive Gaussian noise with standard deviation of 20 followed by bicubic downsampling as a baseline.}
\label{fig:noisecomp}
\end{figure*}

\begin{table}
{\small
\centering
\begin{tabular}{@{}llll@{}}
\toprule
Degradation methods &  & PSNR           & SSIM \\ \midrule
(a) DeResnet~\cite{wei2020unsupervised}  &  & 26.65 & 0.722   \\
(b) Probabilistic DeResnet   &  & \bf{26.96} & \bf{0.744}   \\

\bottomrule
\end{tabular}
\caption{Comparison between deterministic and probabilistic degradation generators.}
\label{tab:abl_noise}
}
\end{table}

\subsection{Collaborative Learning}
Although we can improve the mode coverage by training multiple degradation generators, simply aggregating the generated pseudo-paired data may not be the best way~\cite{zhao2020multi}, implying that there is room for improvement by devising a more clever way to utilize them.
As each data generated by different probabilistic generators shares the content of an image except for the degradation function used, knowledge gained in one can be readily applied to others.
Inspired by He et al.~\cite{he2021multi}, we propose to utilize a collaborative learning strategy, where multiple SR models teach with one another, mutually improving their generalization capabilities.
Figure \ref{fig:overview} shows an overview of the collaborative learning strategy.
With $p_{\theta_i}(\x|\y)$ being the implicit distribution of $G_i$, we train SR networks \(\{\mathcal{M}_{i}\}_{i=1}^K\) by minimizing the objective function

\begin{gather}
    \nonumber \mathcal{L}_i =
    \lambda_{sup}\mathcal{L}_{sup}(\mathcal{M}_i(\x_{i}), \y) \\
    + \lambda_{col}\mathcal{L}_{col}(\{(\mathcal{M}_i(\x_{j}), \mathcal{M}_j(\x_{j}))\}_{j\neq i})\nonumber \\
    + \lambda_{ada}r(p,P)\mathcal{L}_{sup}(\mathcal{M}_i(\x), \hat{\y}),
  \end{gather}
where \(\x_{i}\sim p_{\theta_i}(\x|\y)\), \(\x\sim p(\x)\), and \(\y \sim p(\y)\). \(\lambda_{sup}, \lambda_{col}\), and \(\lambda_{ada}\) denote the weights that determine the relative strength of each term. P and p represent maximum and current training iteration, and thus \(r(p,P) = p/P\) gradually increases during training.
For \(\mathcal L_{sup}\), we used L1-loss, although more complicated loss functions such as adversarial loss or perceptual loss can be plugged in to improve the perceptual quality.
\(\mathcal L_{col}\) is collaborative learning loss, i.e.,

\begin{equation}
    \mathcal{L}_{col}(\{(\mathcal{M}_i(\x_{j}), \mathcal{M}_j(\x_{j}))\}_{j\neq i}) = \\
    \sum _{j\neq i}\mathcal{L}_{sup}(\mathcal{M}_i(\x_{j}), \mathcal{M}_j(\x_{j})).
\end{equation}
Note that knowledge distillation is applied only in image space. We initially tried to apply knowledge distillation in feature space and found that both yield similar results. 
\(\hat{\y}\) is a ensembled pseudo-label for a real LR image \({\x}\), i.e.,
\begin{equation}
    \hat{\y} =  \frac{1}{K}\sum _{l=1}^K \mathcal{M}_l({\x}),
\end{equation}
which helps a model to adapt on the real world dataset.

\section{Experiments}
\subsection{Implementation Details}
\label{sec:imple}
In our experiments, we train our MSSR (Multiple Synthetic data Super Resolution) through two stages. In the first stage, we train two probabilistic degradation generators \(G_1\)  and  \(G_2\) (i.e., $K=2$) based on the code of \cite{wei2020unsupervised}, where we use DeResnet~\cite{wei2020unsupervised} architecture for \(G_1\) and Holistic Attention Network (HAN)~\cite{niu2020single} for \(G_2\).
For the architecture of \(G_2\), we remove the up-sampling block from HAN~\cite{niu2020single}, and strides of the first two Conv-layer are set to 2 for 4x down-sampling.
In the second stage, we train two SR models with RRDBnet~\cite{Wang_2018_ECCV_Workshops} as a SR architecture.
While training SR models, learning rate and batch size are set to 1e-5 and 16, and we set $P$ to 1000000.
Also, we set \(\lambda_{sup}=1\), \(\lambda_{col}=0.01\), and \(\lambda_{ada} = 10\).
We use NTIRE2020 track 1, AIM2019 datasets~\cite{lugmayr2020ntire,lugmayr2019aim} in the remaining parts of this section.

\begin{figure}[t]
	\centering
	\addtolength{\tabcolsep}{-4pt}
	\begin{tabular}{cccc}
	\includegraphics[width=2cm]{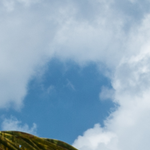}&\includegraphics[width=2cm]{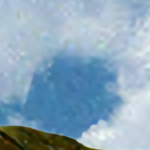}&\includegraphics[width=2cm]{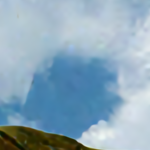}&\includegraphics[width=2cm]{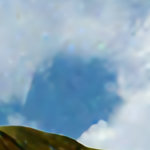}\\
	\includegraphics[width=2cm]{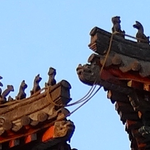}&\includegraphics[width=2cm]{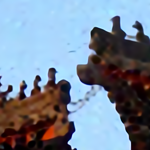}&\includegraphics[width=2cm]{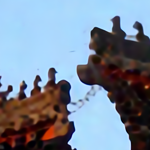}&\includegraphics[width=2cm]{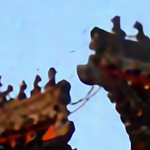}\\
	\includegraphics[width=2cm]{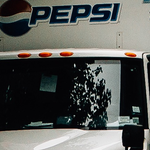}&\includegraphics[width=2cm]{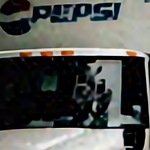}&\includegraphics[width=2cm]{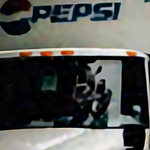}&\includegraphics[width=2cm]{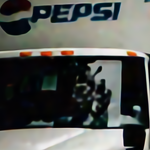}\\
	(a) & (b) & (c) & (d)
	\end{tabular}
	\caption{Qualitative comparison between deterministic and probabilistic degradation generators on NTIRE2020 dataset~\protect\cite{lugmayr2020ntire}. Note that the same SR architecture~\protect\cite{Wang_2018_ECCV_Workshops} is used, and the only difference is degradation models. (a): GT, (b) : DeResnet~\protect\cite{wei2020unsupervised}, (c) : Probabilistic DeResnet, (d) : Probabilistic HAN~\protect\cite{niu2020single}. }
	\label{fig:noiseabsrcomp}
\end{figure}

\begin{table}[]
	
	\centering
	\begin{tabular}{llll}
		
		\multicolumn{2}{c}{}\\
		\toprule
		Methods     & PSNR     & SSIM \\
		\midrule
		Single degradation generator & 26.96 & 0.744 \\
		Naive combination & 27.11  & \bf{0.758}     \\
		CL ($\lambda_{ada} = 0$)& 27.22 & 0.757      \\
		CL & \bf{27.25} & \bf{0.758}  \\
		
		\bottomrule
	
	\end{tabular}
	\caption{Effects of collaborative learning. CL: Collaborative learning. For a single degradation generator setting, we report the results of probabilistic DeResnet in Table~\ref{tab:abl_noise}.  In a naive combination setting, we simply combine the data generated by two probabilistic degradation generators and train a single SR model. For collaborative learning, we report the result of the best performing one out of two models.}
	\label{tab:abl_col}
\end{table}

\

\subsection{ Ablation Studies}

{\bf Probabilistic Degradation Generator} Table \ref{tab:abl_noise} shows the PSNR and SSIM evaluated on NTIRE2020 validation set~\cite{lugmayr2020ntire}. As the results showed, the probabilistic generator significantly improves PSNR from 26.65 to 26.96. Figure \ref{fig:noisecomp} demonstrates that the deterministic generator fails to synthesize the realistic LR image, which is evidenced by high LPIPS. Figure \ref{fig:noiseabsrcomp} shows the qualitative comparison between different degradation methods.
We can see that (b) fails to remove the noise on the LR image as its degradation model is too simple.
On the other hand, (c) and (d) achieves more clear results than (b), demonstrating the superiority of probabilistic degradation generator.
\\

\noindent {\bf Collaborative Learning} Table \ref{tab:abl_col} presents the effect of collaborative learning on NTIRE2020 dataset. Compared to a single degradation generator setting, training multiple degradation generators largely improves the PSNR from 26.96 to 27.11. In addition, the model further improves the PSNR from 27.11 to 27.22 through the collaborative learning. Utilizing pseudo-labels as shown in (b) of Figure \ref{fig:overview} also improves the PSNR from 27.22 to 27.25.

\subsection{Comparison with State-of-the-Arts}

\begin{figure*}[]
	\centering
	\includegraphics[width=13cm]{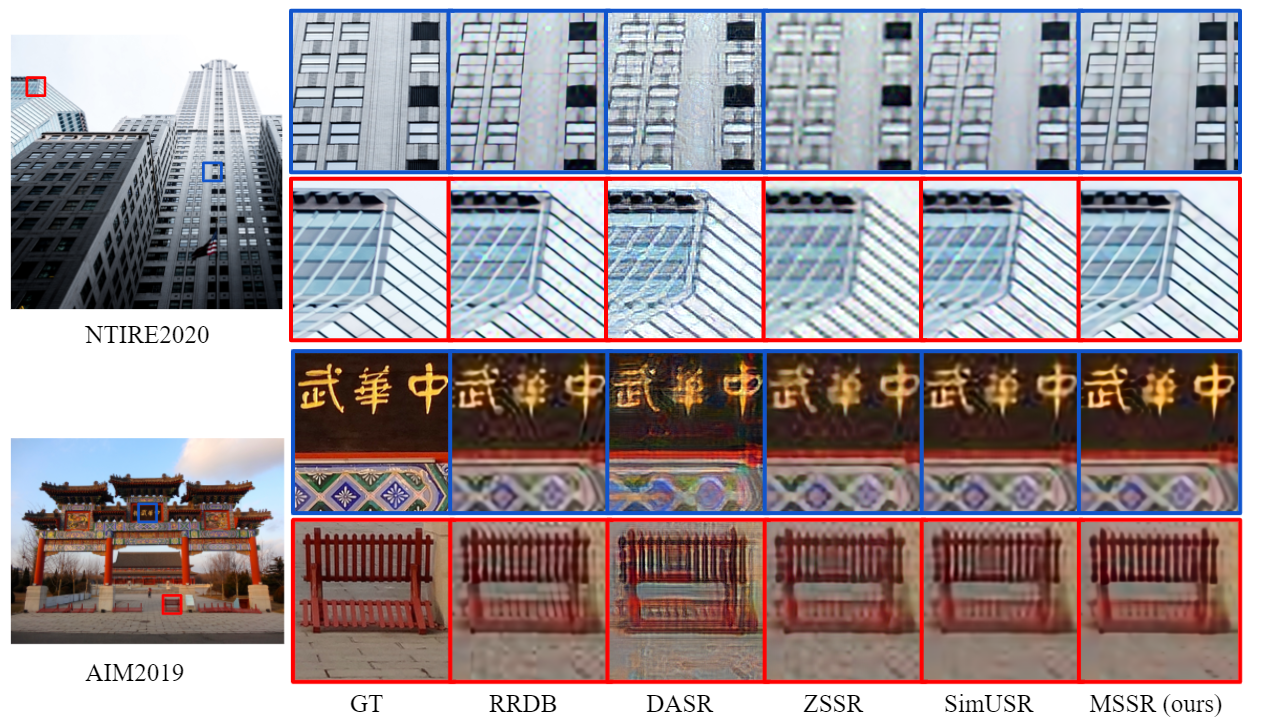}
	\caption{Qualitative comparison with State-of-the-Arts. }
	\label{fig:sota}
\end{figure*}

\begin{table*}[]
\scriptsize
\centering
\begin{tabular}{@{}ccccccccc@{}}
\toprule
Dataset                                                                                    & Metric & CinCGAN & ZSSR*   & Impressionism & FSSR   & DASR   & SimUSR* & MSSR* (ours)     \\ \midrule
\multicolumn{1}{c|}{NTIRE 2020} & PSNR   & 24.19   & 25.33  & 24.96         & 22.41  & 22.95  & 26.48  & \bf{27.25}  \\
\multicolumn{1}{c|}{}                                                                      & SSIM   & 0.683  & 0.662 & 0.664        & 0.519 & 0.491 & 0.725 & \bf{0.758} \\ \midrule
\multicolumn{1}{c|}{AIM 2019} & PSNR   & 21.74        &  22.42      &  21.85             &   20.82     &  21.60      & \bf{22.88}  &   \bf{22.88}              \\
\multicolumn{1}{c|}{}                                                                      & SSIM   &  0.619       &  0.617      &  0.598             &  0.527      & 0.564       & 0.646 &  \bf{0.652}               \\ \bottomrule
\end{tabular}
\caption{Quantitative results of different networks. Results of CinCGAN~\protect\cite{yuan2018unsupervised}, Impressionism~\protect\cite{ji2020real} and FSSR~\protect\cite{fritsche2019frequency} are also reported. * denotes that the method do not use the adversarial loss or perceptual loss to improve the perceptual quality (i.e., PSNR-oriented).}
\label{tab:sotatab}
\end{table*}

Table.~\ref{tab:sotatab} shows that our model achieves the state-of-the-art PSNR and SSIM on the benchmark datasets.
Note that for SimUSR~\cite{ahn2020simusr}, we detach the augmentation~\cite{yoo2020rethinking} used in the original paper for a fair comparison.
As shown in Fig.~\ref{fig:sota}, our model synthesizes more convincing results compared to other methods.

\begin{figure*}[t]
	\centering
    \addtolength{\tabcolsep}{-10pt}
	\begin{tabular}{cc}
	\includegraphics[width=6cm]{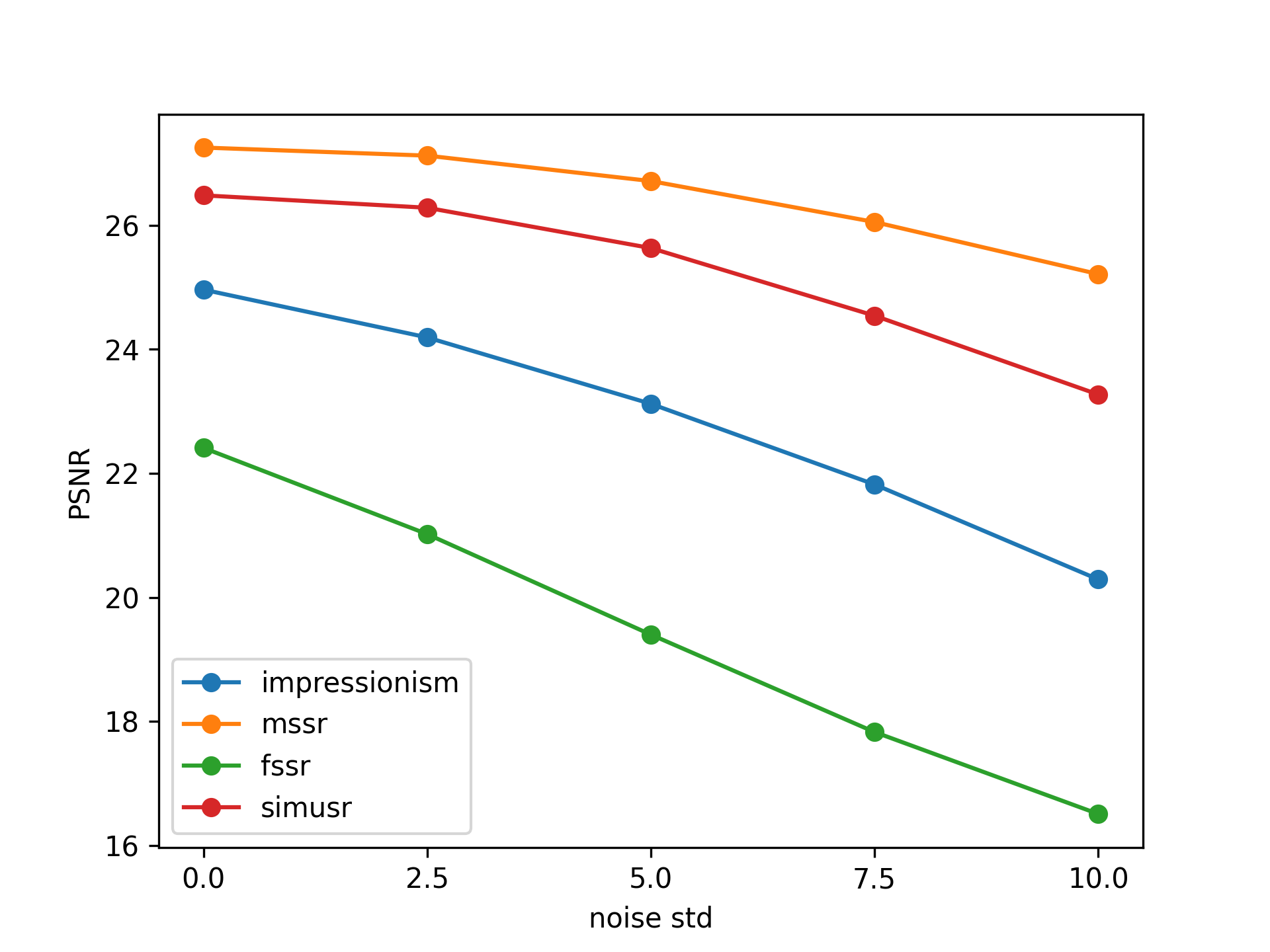} &
	\includegraphics[width=6cm]{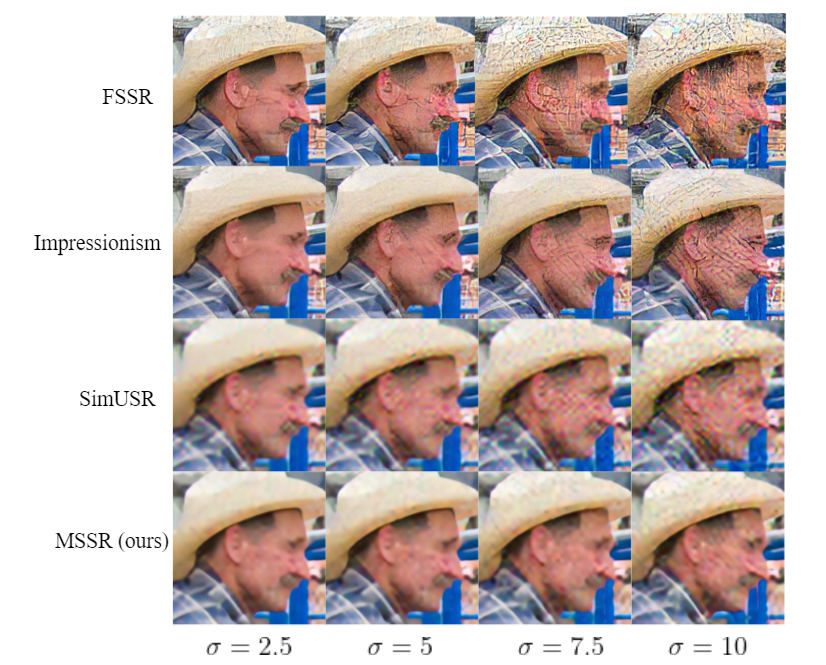} \\
	(a) & (b)

	\end{tabular}
	\caption{Performance degradation on NTIRE2020 dataset due to the increasing magnitude of input perturbation. We inject additive gaussian noise on validation LR images only at the test time to measure the robustness to unseen data.}
	\label{fig:robustness}
\end{figure*}

\subsection{Robustness to Perturbation}
In this section, we construct the experiment to test the generalization capabilities of USR models.
We train SR models on the NTIRE2020 train set and evaluate them on the validation set with additional Gaussian noise perturbation.
As Figure \ref{fig:robustness} indicates, performances of existing methods are drastically degraded by input perturbation, limiting the practical use of SR in the real world. 
In contrast, our model preserves the performance even when the perturbation gets stronger as the probabilistic degradation generators as well as collaborative learning strategy improve the generalization capability of our model to unseen data.

\begin{wrapfigure}[10]{r}{0.45\textwidth}
  \begin{center}
    \includegraphics[width=1\textwidth]{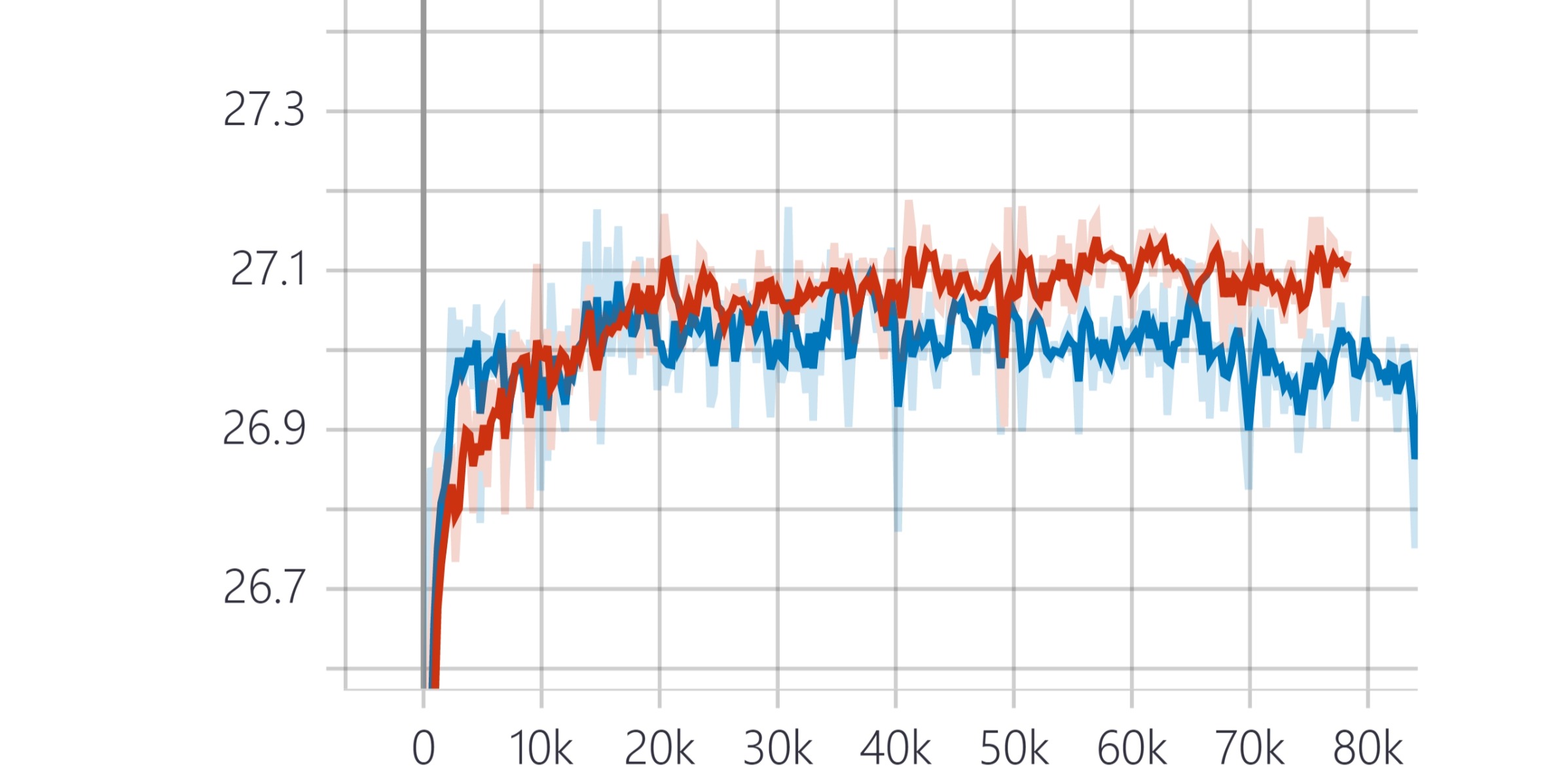}
  \end{center}
  \caption{PSNR of naive combination (blue) and collaborative learning (red) during training.}
  \label{fig:collabo_graph}
\end{wrapfigure}

\subsection{Discussion}
At first glance, collaborative learning appears to be a standard ensemble technique, but it is different from the ensemble as each model exchanges the information via nested knowledge distillation.
We consider that the nested knowledge distillation eases the training procedure by providing rich statistics of $p(\y|\x)$.
Since SR is an ill-posed problem, there are multiple underlying HR representations for a LR observation.
For example, if we use mean-squared loss, it means that a model should learn to predict the expectation of the possible HR images $\mathbb E[p(\y|\x)]$, requiring costly optimization.
Given the limited capacity of a model, it can be problematic as we train a model to invert the multiple degradation generators. 
Contrarily, the proposed collaborative learning enables direct optimization as it can train models with soft labels which contain rich statistics peer models have learned by observing various samples, facilitating the training procedure (see Fig.~\ref{fig:collabo_graph}).

\section{Conclusion}
In this paper, we improve the previous GAN-based USR approaches by training multiple probabilistic degradation generators.
Unlike a deterministic generator, the probabilistic generator can model the stochastic degradation process effectively, improving the generalization capability of SR models.
In addition, we utilize the collaborative learning strategy to adapt to multiple generated subsets effectively.
As a result, MSSR achieved superior results to other methods and showed robustness in unseen data distribution.
These results indicate the effectiveness of our approach for real world SR. 

\section*{Acknowledgement}
This work was supported by Institute of Information \& communications Technology Planning \& Evaluation (IITP) grantfunded by the Korea government(MSIT) (No.2021-0-02068, Artificial Intelligence Innovation Hub).


\clearpage
%
%
\bibliographystyle{splncs04}
\bibliography{reference}
\end{document}